\documentclass[12pt]{article}
\usepackage{epsfig,feynmf}

\makeatletter
\@addtoreset{equation}{section}
\renewcommand{\theequation}{\thesection.\arabic{equation}}
\renewcommand{\thefootnote}{\fnsymbol{footnote}}
\makeatother

\begin{document}
\newcommand{\p}[1]{(\ref{#1})}
\newcommand {\beq}{\begin{eqnarray}}
\newcommand {\eeq}{\end{eqnarray}}
\newcommand {\non}{\nonumber\\}
\newcommand {\eq}[1]{\label {eq.#1}}
\newcommand {\defeq}{\stackrel{\rm def}{=}}
\newcommand {\gto}{\stackrel{g}{\to}}
\newcommand {\hto}{\stackrel{h}{\to}}
\newcommand {\1}[1]{\frac{1}{#1}}
\newcommand {\2}[1]{\frac{i}{#1}}
\newcommand {\thb}{\bar{\theta}}
\newcommand {\ps}{\psi}
\newcommand {\psb}{\bar{\psi}}
\newcommand {\ph}{\varphi}
\newcommand {\phs}[1]{\varphi^{*#1}}
\newcommand {\sig}{\sigma}
\newcommand {\sigb}{\bar{\sigma}}
\newcommand {\Ph}{\Phi}
\newcommand {\Phd}{\Phi^{\dagger}}
\newcommand {\Sig}{\Sigma}
\newcommand {\Phm}{{\mit\Phi}}
\newcommand {\eps}{\varepsilon}
\newcommand {\del}{\partial}
\newcommand {\dagg}{^{\dagger}}
\newcommand {\pri}{^{\prime}}
\newcommand {\prip}{^{\prime\prime}}
\newcommand {\pripp}{^{\prime\prime\prime}}
\newcommand {\prippp}{^{\prime\prime\prime\prime}}
\newcommand {\pripppp}{^{\prime\prime\prime\prime\prime}}
\newcommand {\delb}{\bar{\partial}}
\newcommand {\zb}{\bar{z}}
\newcommand {\mub}{\bar{\mu}}
\newcommand {\nub}{\bar{\nu}}
\newcommand {\lam}{\lambda}
\newcommand {\lamb}{\bar{\lambda}}
\newcommand {\kap}{\kappa}
\newcommand {\kapb}{\bar{\kappa}}
\newcommand {\xib}{\bar{\xi}}
\newcommand {\ep}{\epsilon}
\newcommand {\epb}{\bar{\epsilon}}
\newcommand {\Ga}{\Gamma}
\newcommand {\rhob}{\bar{\rho}}
\newcommand {\etab}{\bar{\eta}}
\newcommand {\chib}{\bar{\chi}}
\newcommand {\tht}{\tilde{\th}}
\newcommand {\zbasis}[1]{\del/\del z^{#1}}
\newcommand {\zbbasis}[1]{\del/\del \bar{z}^{#1}}
\newcommand {\vecv}{\vec{v}^{\, \prime}}
\newcommand {\vecvd}{\vec{v}^{\, \prime \dagger}}
\newcommand {\vecvs}{\vec{v}^{\, \prime *}}
\newcommand {\alpht}{\tilde{\alpha}}
\newcommand {\xipd}{\xi^{\prime\dagger}}
\newcommand {\pris}{^{\prime *}}
\newcommand {\prid}{^{\prime \dagger}}
\newcommand {\Jto}{\stackrel{J}{\to}}
\newcommand {\vprid}{v^{\prime 2}}
\newcommand {\vpriq}{v^{\prime 4}}
\newcommand {\vt}{\tilde{v}}
\newcommand {\vecvt}{\vec{\tilde{v}}}
\newcommand {\vecpht}{\vec{\tilde{\phi}}}
\newcommand {\pht}{\tilde{\phi}}
\newcommand {\goto}{\stackrel{g_0}{\to}}
\newcommand {\tr}{{\rm tr}\,}
\newcommand {\GC}{G^{\bf C}}
\newcommand {\HC}{H^{\bf C}}
\newcommand{\vs}[1]{\vspace{#1 mm}}
\newcommand{\hs}[1]{\hspace{#1 mm}}
\newcommand{\al}{\alpha}
\newcommand{\be}{\beta}
\newcommand{\Lam}{\Lambda}
\newcommand{\kahler}{K\"ahler }
\newcommand{\con}[1]{{\Gamma^{#1}}}
\newcommand{\sect}[1]{\setcounter{equation}{0}\section{#1}}
\renewcommand{\theequation}{\thesection.\arabic{equation}}

\thispagestyle{empty}
\begin{flushright}
{\tt HIP-2007-01/TH \\
\tt KEK-TH-1128\\
\tt hep-ph/0701155} \\
\end{flushright}
\begin{center}
{\Large
{\bf Top quark spin correlations 
 in the Randall-Sundrum model 
 at the CERN Large Hadron Collider 
}}
\\[5mm]

\normalsize
{\large \bf
  Masato~Arai~$^{a}$}
\footnote{\it
masato.arai@helsinki.fi}
,
{\large \bf
  Nobuchika~Okada~$^{b}$}
\footnote{\it
okadan@post.kek.jp}
,
{\large \bf Karel Smolek~$^{c}$}
\footnote{\it
karel.smolek@utef.cvut.cz
}
\\
and 
\\
{\large \bf
Vladislav \v{S}im\'ak~$^{d}$}
\footnote{\it
simak@fzu.cz
}

\vskip 1.0em

{$^{a}$ \it High Energy Physics Division, 
             Department of Physical Sciences,
             University of Helsinki \\
 and Helsinki Institute of Physics,
 P.O.Box 64, FIN-00014, Finland
         \\
 $^{b}$ Theory Division, KEK, Tsukuba, 
        Ibaraki 305-0801, Japan \\
 $^{c}$ \it Institute of Experimental and Applied Physics, \\
        Czech Technical University in Prague, 
        Horsk\'a 3a/22, 128 00 Prague 2, Czech Republic \\
 $^{d}$ \it  Faculty of Nuclear Sciences and Physical Engineering,\\
        Czech Technical University in Prague, 
        B\v{r}ehov\'a 7, 115 19 Prague 1, Czech Republic
}
\vspace{3mm}
{\bf Abstract}\\[5mm]
{\parbox{16cm}{
In the Randall-Sundrum model, 
 we study top-antitop pair production 
 and top spin correlations at the Large Hadron Collider. 
In addition to the Standard Model processes, 
 there is a new contribution to the top-antitop pair production 
 process mediated by graviton Kaluza-Klein modes in the $s$-channel. 
We calculate the density matrix for the top-antitop pair 
 production including the new contribution. 
With a reasonable parameter choice in the Randall-Sundrum model, 
 we find a sizable deviation of the top-antitop pair production 
 cross section and the top spin correlations 
 from those in the Standard Model. 
In particular, resonant productions of 
 the graviton Kaluza-Klein modes 
 give rise to a remarkable enhancement of such a deviation. 
}}
\end{center}
\vfill
\newpage
\setcounter{page}{1}
\setcounter{footnote}{0}
\renewcommand{\thefootnote}{\arabic{footnote}}
%
%
\section{Introduction}
During the past several decades the gauge hierarchy problem 
 has been a guiding principle to propose 
 beyond the standard model (SM), and  
 many new physics models have been proposed to solve this problem. 
Brane world scenario recently proposed provides a possible solution 
 for this problem. 
In this scenario whole space has more than three spatial dimensions 
 and the SM fields are confined on a 4-dimensional hypersurface 
 called ``D3-brane''. 
There are two typical models based on this setup. 
One is the so-called ADD model proposed by 
 Arkani-Hamed, Dimopoulos and Dvali (ADD) \cite{ADD}.
In this model, there are $n$-extra dimensions compactified 
 on $n$-torus with common radius $R$ and a D3-brane 
 embedded in $(4+n)$-dimensional bulk is introduced 
 on which the SM fields reside. 
This setup gives a relation $M_{\rm pl}=M_{\rm D}(M_{\rm D} R)^{n/2}$ 
 between the 4-dimensional Planck mass $M_{\rm pl}$ 
 and the Planck scale of $(4+n)$-dimensions $M_{\rm D}$. 
If the compactification radius is large enough 
 (for instance, $R\sim 0.1$ mm for $n=2$), 
 $M_D$ can be ${\cal O}$(1 TeV) and thus one obtains 
 a solution to the gauge hierarchy problem. 
In fact, this picture is consistent with the current experimental 
 bound on $R$ around $200$ $\mu$m \cite{upper}.

The other model was proposed by Randall and Sundrum (RS) \cite{RS}. 
This is a 5-dimensional model, 
 where one extra-dimension is compactified 
 on a ${{\bf S}^1 /Z_2}$ orbifold and 
 a negative cosmological constant is introduced in the bulk. 
Two D3-branes are placed at fixed points 
 of the orbifold $\phi=0$ and $\phi=\pi$ 
 ($\phi$ is an angle of ${\bf S}^1$) with opposite brane tensions. 
A brane at $\phi=0$ with a positive tension is called 
 the hidden brane and the other one at $\phi=\pi$ 
 with a negative tension is called the visible brane 
 on which the SM fields are confined. 
Solving the Einstein equation of this system, 
 the 5-dimensional bulk geometry is found to be a slice 
 of anti-de Sitter (AdS$_5$) space, 
\begin{eqnarray}
 ds^2=e^{-2 \kappa r_c |\phi|} \eta_{\mu\nu}dx^\mu dx^\nu - r_c^2 d\phi^2\, , 
~~\eta_{\mu\nu}={\rm diag}(1,-1,-1,-1)\,,
 \label{geometry}
\end{eqnarray}
where $\kappa$ is the AdS curvature in five dimensions, 
 and $r_c$ is a compactification radius. 
This background geometry allows us to take the Planck scale 
 as a fundamental scale. 
Indeed, in effective 4-dimensional description 
  an effective mass scale on the visible brane 
 is warped down such as 
 $\Lambda_\pi = \bar{M}_{\rm pl}e^{-\pi\kappa r_c}$ 
 due to effect of the warped geometry, 
 where $\bar{M}_{\rm pl}$ is the reduced Planck mass. 
Therefore, with a mild parameter tuning, $\kappa r_c\simeq 12$, 
 we can realize  $\Lambda_\pi = {\cal O}$(1 TeV) 
 and obtain a natural solution to the gauge hierarchy problem.

In the brane world scenario, an infinite tower of Kaluza-Klein (KK) gravitons  
 appears in effective 4-dimensional theory. 
Effective couplings between these KK gravitons and the SM fields  
 are controlled by $M_D$ or $\Lambda_\pi$ in each typical model. 
Since these mass scales should be around TeV 
 so as to solve the gauge hierarchy problem, 
 we can expect new phenomena induced by the KK gravitons, 
 for example, 
 direct KK graviton emission process and virtual KK graviton 
 exchange process at high energy collisions. 
In particular, the virtual KK graviton exchange process 
 is interesting, because it can give rise to characteristic 
 angular distributions and spin configurations 
 for outgoing particles, which reflect the spin-2 nature 
 of the intermediate KK gravitons.

One of good candidates to study a spin configuration 
 is a top-antitop quark pair, 
  since the top quark, with mass in the range of $175$ GeV \cite{Abe}, 
  decays electroweakly before hadronizing \cite{Bigi}.
A possible spin polarization of the top-antitop quark pair 
 is directly transferred to its decay products and therefore 
 there are significant angular correlations 
 between the top quark spin axis and the direction of motion 
 of the decay products. 
The spin correlations for the hadronic top-antitop pair production 
 process have been extensively studied in the quantum 
 chromodynamics (QCD) \cite{Stelzer, Mahlon-Parke, Bernreuther2}. 
It is found that there is a spin asymmetry between 
 the produced top-antitop pairs, namely, 
 the number of produced top-antitop quark pairs with both spin up or 
 spin down (like pair) is different from the number of pairs 
 with the opposite spin combinations (unlike pair). 
If the top quark is coupled to a new physics beyond the SM, 
 the top-antitop spin correlations could be altered. 
Therefore, the top-antitop spin correlations can provide 
 useful information to test not only the SM but also 
 a possible new physics at hadron colliders. 
The Large Hadron Collider (LHC) has a big advantage 
 to study the top spin correlations, 
 since it will produce almost 10 millions of top quarks a year 
 (during its low luminosity run).

In Ref. \cite{AOSS}, 
 effect of the KK gravitons on the top spin correlations 
 in the ADD model at the LHC was studied. 
A sizable deviation of the top spin correlations 
 from the SM one was found with 
 scale $M_D$ below 2 TeV. 
The purpose of this paper is to study the top spin correlations 
 in the RS model at the LHC. 
To study this issue in the RS model is more motivated than 
 in the ADD model by the following reasons. 
In the ADD model, a mass difference of each KK graviton 
 is characterized by the radius of the extra dimensions 
 ($R^{-1} \sim $meV for $n=2$), 
 which is much smaller than detector resolutions 
 and it is impossible to identify each resonant KK graviton 
 at collider experiments. 
In fact, couplings between each KK graviton and the SM fields 
 are suppressed by the 4-dimensional Planck mass and extremely weak. 
After coherently summing up many KK graviton processes, 
 the KK graviton effects can be sizable. 
However, there is a theoretical problem 
 in the ADD model with two or more extra dimensions: 
 Sum of all intermediate KK gravitons diverges and 
 is not well-defined. 
Although this problem can be solved by introducing a finite brane 
 tension \cite{Bando:1999di} 
 or a finite brane width \cite{Hisano:1999bn}, 
 which give rise to a physical ultraviolet cutoff and make 
 the sum finite, 
 a new parameter (the brane tension or the width of the brane) 
 is brought into a model. 
On the other hand, in the RS model 
 only one extra dimension is introduced, and 
 sum of all intermediate KK gravitons turns out to be finite 
 and the KK graviton mediated process is well-defined 
 at low energies. 
Each KK graviton strongly couples to the SM fields 
 with $\Lambda_\pi$ suppressed couplings, 
 and KK graviton mass is characterized by 
 $\kappa e^{- \kappa r_c \pi} \sim$ TeV. 
As a result, we can expect a resonant production of the KK gravitons 
 at colliders if the collider energy is high enough. 
This is a direct signal of the RS model. 
Furthermore, the resonance gives rise to an enhancement 
 of production of the top-antitop pairs 
 and provides a big statistical advantage 
 for studying the top spin correlations around the resonance pole.

This paper is organized as follows.
In section 2, we briefly review the top spin correlations. 
In section 3, we examine the invariant amplitudes for the polarized 
top-antitop pair production processes, $q\bar{q}\rightarrow t\bar{t}$,
 $gg\rightarrow t\bar{t}$ mediated by virtual KK gravitons in the
 $s$-channel. 
We perform numerical analysis in section 4. 
Section 5 is devoted to conclusions.
In Appendices, the coefficient of the KK graviton decay width and the
 density matrix are shown.

\section{Spin correlation}
At hadron collider, the top-antitop quark pair is produced 
 through the processes of quark-antiquark pair annihilation 
 and gluon fusion:
\begin{eqnarray}
 &i \rightarrow t+\bar{t}, \,\,\, i=q\bar{q}\,,gg\,.& \label{top1}
\end{eqnarray}
The former is the dominant process at the Tevatron, 
 while the latter is dominant at the LHC. 
The produced top-antitop pairs decay before hadronization takes place. 
The main decay modes in the SM 
 involve leptonic and hadronic modes: 
\begin{eqnarray}
 &t\rightarrow bW^+ \rightarrow bl^+\nu_l\,,bu\bar{d}\,,bc\bar{s},&
 \label{decay}
\end{eqnarray}
where $l=e,\mu,\tau$.
The differential decay rates to a decay product $f=b,l^+, \nu_l,$ etc.  
 at the top quark rest frame can be parameterized as 
\begin{eqnarray}
{1 \over \Gamma}{d \Gamma \over d \cos \theta_f}=
  {1 \over 2}(1 + \kappa_f \cos \theta_f ), 
 \label{decay1}
\end{eqnarray}
where $\Gamma$ is the partial decay width of 
 the respective decay channel and 
 $\theta_f$ is the angle between the 
 top quark polarization
 and the direction of motion of the decay product $f$
 at the top quark rest frame.
The coefficient $\kappa_f$ called top spin analyzing power 
 is a constant between $-1$ and $1$.
The ability to distinguish 
 the polarization
 of the top quark evidently increases with $\kappa_f$.
The most powerful spin analyzer is a charged lepton, for which 
 $\kappa_{l^+}=+1$ at tree level \cite{Jezabek}. 
Other values of $\kappa_f$ are
 $\kappa_b = -0.41$ for the $b$-quark 
 and $\kappa_{\nu_l}=-0.31$ for the $\nu_l$, respectively. 
In hadronic decay modes, the role of the charged lepton 
 is replaced by the $d$ or $s$ quark.

Now we see how the top spin correlations appear in the chain of processes 
 of $i\rightarrow t\bar{t}$ and decay of the top quarks.
The total matrix element squared 
 for the top-antitop pair production \p{top1} 
 and their decay channels \p{decay} is given by 
\begin{eqnarray}
|{\cal M}|^2 \propto {\rm Tr}[\rho R^i \bar{\rho}]
 =\rho_{\alpha^\prime\alpha}R^i_{\alpha\beta,\alpha^\prime\beta^\prime}
  \bar{\rho}_{\beta^\prime\beta} \label{comp}
\end{eqnarray}
in the narrow-width approximation for the top quark.
Here the subscripts denote the top and antitop spin indices, 
 and $R^i$ denotes the density matrix 
 corresponding to the production of the on-shell top-antitop quark pair 
 through the process $i$ in \p{top1}:
\begin{eqnarray}
 R_{\alpha\beta,\alpha^\prime\beta^\prime}^i
 =\sum_{{\rm initial~spin}}{\cal M}(i\rightarrow t_\alpha\bar{t}_\beta)
  {\cal M}^*(i\rightarrow t_{\alpha^\prime}\bar{t}_{\beta^\prime}),
\end{eqnarray}
where ${\cal M}(i\rightarrow t_\alpha\bar{t}_\beta)$ is 
 the amplitude for the top-antitop pair production. 
The matrices $\rho$ and $\bar{\rho}$ are the density matrices 
 corresponding to the decays of polarized top and antitop quarks 
 into some final states at the top and antitop rest frame, respectively.
In the leptonic decay modes, 
 the matrices $\rho$, which lead to \p{decay1},
 can be obtained as (see, for instance, \cite{Bernreuther})
\begin{eqnarray}
  \rho_{\alpha^\prime\alpha}
  = {\cal M}(t_\alpha \rightarrow bl^+\nu_l)
    {\cal M}^*(t_{\alpha^\prime} \rightarrow bl^+\nu_l)  
  = {\Gamma \over 2}(1 + \kappa_f {\vec{\sigma}} \cdot 
       \vec{q}_f)_{\alpha^\prime\alpha},  
 \label{rho1}
\end{eqnarray}
where $q_f$ is the unit vector of the direction of motion 
 of the decay product $f$. 
The density matrix for the polarized antitop quark 
 is obtained by replacing $\kappa_f \rightarrow -\kappa_f$ in \p{rho1}
 if there is no CP violation.
In the SM, there is no CP violation in the top quark decay at the
 leading order. 
In the RS model we consider, there is no CP violation 
 at the leading order, and this relation holds. 

The best way to analyze the top-antitop spin correlations 
 is to see the angular correlations 
 of two charged leptons $l^+l^-$ 
 produced by the top-antitop quark leptonic decays. 
In the following, we consider only the leptonic decay channels. 
Using \p{comp}-\p{rho1} and integrating over 
 the azimuthal angles of the charged leptons, 
 we obtain the following double distribution
 \cite{Stelzer, Mahlon-Parke, Bernreuther2}
\begin{eqnarray}
 {1 \over \sigma}
 {d^2 \sigma \over d \cos\theta_{l^+} d \cos\theta_{l^-}}
= {1 \over 4}\left({1+B_1 \cos\theta_{l^+}
 +B_2 \cos\theta_{l^-}-C
    \cos\theta_{l^+} \cos\theta_{l^-}}\right).
 \label{double}
\end{eqnarray}
Here $\sigma$ denotes the cross section 
 for the process of the leptonic decay modes, 
 and $\theta_{l^+} (\theta_{l^-})$ denotes the angle 
 between the top (antitop) spin axis and 
 the direction of motion of the antilepton (lepton) 
 at the top (antitop) rest frame. 
In the following analysis, we use the helicity spin basis which is almost optimal
 one to analyze the top spin correlation at the LHC.\footnote{Recently another
 spin basis was constructed, which has a larger spin correlation than the
 helicity basis at the LHC \cite{uwer}.}
In this basis, the top (antitop)
 spin axis is regarded as the direction of motion of the top (antitop) in
 the top-antitop center-of-mass system.
The coefficients $B_1$ and $B_2$ are associated with 
 a possible polarization of the (anti)top quark transverse to the
 production plane in proton-proton (or proton-antiproton) collision, 
 called a transverse polarization,
 and $C$ encodes the top spin correlations,
 whose explicit expression is given by
\begin{eqnarray}
 C= {\cal A} \kappa_{l^+}\kappa_{l^-},~~~\kappa_{l^+}=\kappa_{l^-}=1\,,
\end{eqnarray}
where the coefficient ${\cal A}$ represents the spin asymmetry 
 between the produced top-antitop pairs 
 with like and unlike spin pairs defined as 
\begin{eqnarray}
 {\cal A}={\sigma(t_\uparrow\bar{t}_\uparrow)
          +\sigma(t_\downarrow\bar{t}_\downarrow) 
          -\sigma(t_\uparrow\bar{t}_\downarrow)
          -\sigma(t_\downarrow\bar{t}_\uparrow) 
         \over
           \sigma(t_\uparrow\bar{t}_\uparrow)
          +\sigma(t_\downarrow\bar{t}_\downarrow)
          +\sigma(t_\uparrow\bar{t}_\downarrow)
          +\sigma(t_\downarrow\bar{t}_\uparrow)}.
\label{asym}
\end{eqnarray}
Here $\sigma(t_\alpha\bar{t}_\beta)$ is 
 the cross section of the top-antitop pair production 
 at parton level with denoted spin indices.

In the SM, there is no transverse polarization, $B_1=B_2=0$ at the leading
 order of $\alpha_s$\footnote{At the one-loop level, the transverse
 polarization is induced. Detailed analysis has been performed 
 in Refs. \cite{Dharmaratna:1989jr} and \cite{Bernreuther:1995cx}.} while
 the spin asymmetry is found to be ${\cal A}=+0.319$ 
 for the LHC. \footnote{
The parton distribution function set of CTEQ6L \cite{CTEQ} 
 has been used in our calculations. 
The resultant spin asymmetry somewhat depends 
 on the parton distribution functions used. } 
At the LHC in the ATLAS experiment, the spin asymmetry 
 of the top-antitop pairs will be 
 measured with a precision of several percent, even after one LHC year 
 at low luminosity (10 fb$^{-1}$) \cite{ATLAS_TOP}.
Since in the brane world scenario 
 there is a new contribution to the top-antitop 
 production process through virtual KK graviton exchange 
 in the $s$-channel, 
 the spin asymmetry could be altered from the SM one. 
It is found that in the ADD model, 
 the KK graviton contribution reduces the spin asymmetry 
 \cite{AOSS}, for example, ${\cal A}=+0.147$ for $M_D=1$ TeV. 
In the next section, we calculate the squared amplitude 
 for the top-antitop pair production 
 including the virtual KK graviton mediated process 
 in the RS model and evaluate the spin asymmetry at tree level,
 assuming that loop corrections in the RS model will not give 
 a significant contribution.

\section{Scattering amplitude in the RS model}

In the RS model, because of the warped metric, 
 zero-mode graviton and KK gravitons 
 have different non-trivial configurations 
 with respect to the fifth dimensional coordinates. 
In particular, the KK gravitons are localizing 
 around the visible brane and so couplings 
 between the KK gravitons and the SM fields are enhanced. 
The effective interaction Lagrangian is given by \cite{rizzo}
\begin{eqnarray}
{\cal L}_{\rm int}=-{1 \over \bar{M}_{\rm pl}}
 T^{\mu\nu}(x)h_{\mu\nu}^{(0)}(x) 
 -{1 \over \Lambda_\pi}
 T^{\mu\nu}(x)\sum_{n=1}^\infty
 h_{\mu\nu}^{(n)}(x)\,, \label{int-L}
\end{eqnarray}
where 
 $h_{\mu\nu}^{(n)}$ is the $n$-th graviton KK mode, 
 $T^{\mu\nu}$ is the energy-momentum tensor 
 of the SM fields on the visible brane, 
 and 
$\Lambda_{\pi}= \bar{M}_{\rm pl} e^{-\kappa r_c \pi} \sim$ TeV. 
The graviton zero mode couples with the usual strength 
 and its effect is of course negligible for collider physics, 
 while each graviton KK mode strongly couples to the SM fields 
 with the suppression factor $\Lambda_{\pi}^{-1}$. 

Mass spectrum of the KK gravitons is determined by the relation 
\begin{eqnarray}
 m_n=x_n \kappa e^{-\kappa r_c \pi}\,,
\end{eqnarray}
where $x_n$ is a root of the Bessel function of the first order, 
 $J_1(x_n)=0$, and 
 $ x_1\sim 3.83$, $x_2\sim 7.02$, $x_3\sim 10.17$, for example. 
Assuming that the 5-dimensional curvature $\kappa$ 
 is small compared to 
 $M$ where $M$ is the 5-dimensional Planck scale,
 the lightest KK graviton mass appears around several hundred GeV 
 which is accessible by the LHC. 
Once the lightest KK graviton mass is fixed, 
 higher KK graviton mass can be determined by using 
 given numerical factors, $m_n= m_1 (x_n/x_1)$.  
Using $m_1$, the effective scale $\Lambda_\pi$ can be rewritten as 
\begin{eqnarray}
 \Lambda_\pi = 
  \frac{m_1}{3.83} \left( 
  \frac{\bar{M}_{\rm pl}}{\kappa} 
  \right). \label{coupling}
\end{eqnarray}
In our numerical analysis, we use 
 $m_1$ and $\kappa/\bar{M}_{\rm pl}$ as input parameters. 
As mentioned above, we assume the 5-dimensional curvature $\kappa$ is
 much smaller than $M$, whose
 condition is actually necessary to trust the RS metric.
It yields the bound for the input parameter as
 $\kappa/\bar{M}_{\rm pl}<0.1$ \cite{rizzo}.

The effective interaction Eq. \p{int-L} 
 leads to the top-antitop pair production 
 through the virtual KK graviton exchange in the $s$-channel. 
The invariant amplitude in momentum space 
 is obtained as\footnote{
This is easily reproduced by the replacement 
 in a corresponding formula in the ADD model: 
 ${4\pi\lambda \over  M_D^4} 
 \rightarrow -{1 \over \Lambda_\pi^2}\sum_{n=1}^\infty{1 \over s-m_n^2}$
 in the normalization of Ref. \cite{AOSS}.}
\begin{eqnarray}
{\cal M}_G 
 &=&A(s)T^{\mu \nu} (k_1 , k_2) T_{\mu \nu} (k_3 , k_4)\,, \\
 A(s)&\equiv &-{1 \over \Lambda_\pi^2}
      \sum_{n=1}^\infty \frac{1}{s^2- m_{n}^2+im_n \Gamma_n}\, , 
 \label{a} 
\end{eqnarray}
with $s=(k_1+k_2)^2=(k_3+k_4)^2$.
Here $\Gamma_n$ is the total decay width of the $n$-th KK graviton 
 given by 
\begin{eqnarray}
 \Gamma_n(h^{(n)}\rightarrow yy)
 ={m_n x_n^2 \over 16 \pi}\left(
 {\kappa \over \bar{M}_{\rm pl}}\right)^2
 \sum_{y}\Delta_n^{yy}\,,
\end{eqnarray}
where $\Delta_n^{yy}$ is a dimensionless coefficient 
 for each decay mode, and its exact form is listed in Appendix \ref{coef}. 

In the center-of-mass frame, we can straightforwardly calculate 
 the density matrix $R^i_{\alpha\beta,\alpha^\prime\beta^\prime}$ 
 including both the SM (QCD) and the virtual KK-graviton contributions.
It is sufficient to calculate its diagonal part 
 since only the diagonal part is relevant to the spin asymmetry
 \p{asym} (full density matrix is shown in Appendix \ref{matrix}).
For the $q\bar{q}$ initial state we find 
\begin{eqnarray}
 |{\cal M}(q\bar{q}\rightarrow t_\uparrow\bar{t}_\uparrow)|^2
  &=& |{\cal M}(q\bar{q}\rightarrow {t_\downarrow\bar{t}_\downarrow})|^2 \non
  &=& {g^4 \over 9}(1-\beta^2)\sin^2\theta
  +{|A(s)|^2 s^4\beta^2 \over 128}(1-\beta^2)\sin^22\theta\,, 
  \label{qq1} \\
 |{\cal M}(q\bar{q}\rightarrow t_\uparrow\bar{t}_\downarrow)|^2
 &=& |{\cal M}(q\bar{q}\rightarrow {t_\downarrow\bar{t}_\uparrow})|^2 \non
  &=& {g^4 \over 9}(1+\cos^2\theta)+{|A(s)|^2s^4\beta^2 \over 128}
      (\cos^22\theta+\cos^2\theta)\,,
  \label{qq2}
\end{eqnarray}
where 
 $\theta $ is the scattering angle between the incoming $q$ 
 and outgoing $t$, 
 $g$ is a strong coupling constant,  
 and  $\beta=\sqrt{1-4m^2_t/s}$ with top quark mass $m_t$. 
For the $gg$ initial state we obtain
\begin{eqnarray} 
 &&|{\cal M}(gg \rightarrow t_\uparrow\bar{t}_\uparrow)|^2
    = |{\cal M}(gg \rightarrow {t_\downarrow\bar{t}_\downarrow})|^2 \non 
 &&~~~~
    ={g^4 \beta^2 \over 96}{\cal Y}(\beta,\cos\theta)(1-\beta^2)
     (1+\beta^2+\beta^2\sin^4\theta)+{\cal Z}(\beta,\theta,s)
     s^2 \beta^2(1-\beta^2)\sin^4\theta\,, \non \label{gg1} \\
 &&|{\cal M}(gg\rightarrow t_\uparrow\bar{t}_\downarrow)|^2
   = |{\cal M}(q\bar{q}\rightarrow {t_\downarrow\bar{t}_\uparrow})|^2 \non
 &&~~~~
     ={g^4 \beta^2 \over 96}{\cal Y}(\beta,\cos\theta)\sin^2\theta(1+\cos^2\theta)
      +{\cal Z}(\beta,\theta,s)
       s^2 \beta^2\sin^2\theta(1+\cos^2\theta)\,. \label{gg2}
\end{eqnarray}
Here $ {\cal Y}(\beta,\theta,s)$ and $ {\cal Z}(\beta,\theta,s)$ 
 are defined by
\begin{eqnarray}
 {\cal Y}(\beta,\cos\theta)&=&
  {7+9\beta^2\cos^2 \theta \over (1-\beta^2\cos^2\theta)^2}\,,~~
  \label{funcs1} \\
 {\cal Z}(\beta,\theta,s)&=&{1 \over 32}
   \left(
   -{g^2 \over 1-\beta^2\cos^2\theta}{\rm Re}(A(s))+{3 \over 8}|A(s)|^2 s^2
   \right)\,, \label{funcs2}
\end{eqnarray}
respectively.
As in the same with the ADD case discussed in Ref. \cite{AOSS}, 
 there is no interference term for the quark-antiquark pair annihilation process, 
 while there is the non-vanishing interference in the gluon fusion process.

With the squared amplitudes Eqs. \p{qq1}-\p{gg2},
 one can find the integrated top-antitop quark pair production 
 cross section through the formula, 
\begin{eqnarray}
 \sigma_{tot}( pp \rightarrow t_\alpha\bar{t}_{\beta})&=&
  \sum_{a,b} \int dx_1 \int dx_2 \int d\cos\theta 
   f_a(x_1,Q^2)f_b(x_2,Q^2) \nonumber \\
&&~~~~~~~\times{d \sigma(a(x_1E_{\rm CMS}/2)b(x_2E_{\rm CMS}/2)\rightarrow 
   t_\alpha\bar{t}_{\beta}) \over d \cos\theta} \,,
 \label{total}
\end{eqnarray} 
where $f_a$ denotes the parton distribution function 
 for a parton $a$,  
 $E_{\rm CMS}$ is a center-of-mass energy of a proton-proton system,
 and $Q$ is 
a factorization scale.

Using above formulas, we calculate the double distribution
 (\ref{double}) in the RS model.
Explicit calculation 
 tells us that the transverse polarization is
 vanishing, i.e. $B_1=B_2=0$ in the RS model while the spin asymmetry
 ${\cal A}$ is altered from the SM one.

\section{Numerical results}

Here we show various numerical results 
 and demonstrate interesting properties 
 of measurable quantities in the RS model. 
In our analysis we use the parton distribution functions of 
 CTEQ6L \cite{CTEQ} with the 
 factorization scale $Q=m_t=175$
 GeV, $N_f=5$ and $\alpha_s(Q)=0.1074$.  
As mentioned above, we choose $m_1$ and $\kappa/\bar{M}_{\rm pl}$ 
 as input parameters. 
In practice, we fix $m_1=600$ GeV, 
 subsequently $m_2=1099$, $m_3=1582$, $m_4=2686$ GeV etc. 

\begin{figure}[H]
\begin{center}
  \epsfxsize=8.2cm
  \epsfbox{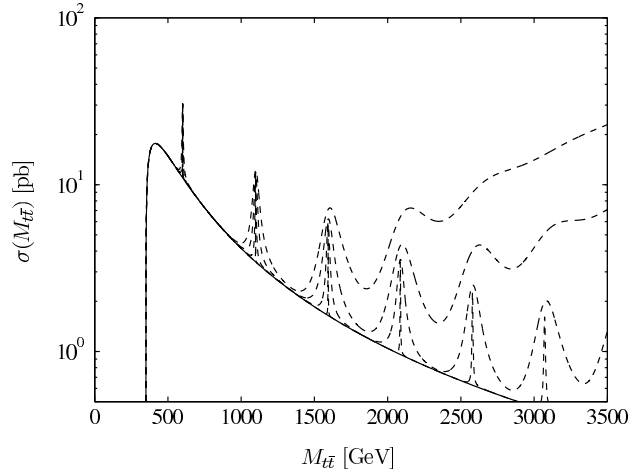}
\caption{
 The dependence of the 
 cross section of the top-antitop 
 quark pair production by quark-antiquark pair annihilation 
 on the center-of-mass energy of colliding partons. 
The solid line and dashed lines correspond to 
 the results of the SM and the RS model for $m_1=600$ GeV 
 and $\kappa/{\bar{M}_{\rm pl}}=0.01, 0.04, 0.07$ 
 and $0.1$ from bottom to top, respectively. 
}
  \label{fig_sigma_omega_qq}
\end{center}
\end{figure}

\begin{figure}[H]
\begin{center}
  \epsfxsize=8.2cm
  \epsfbox{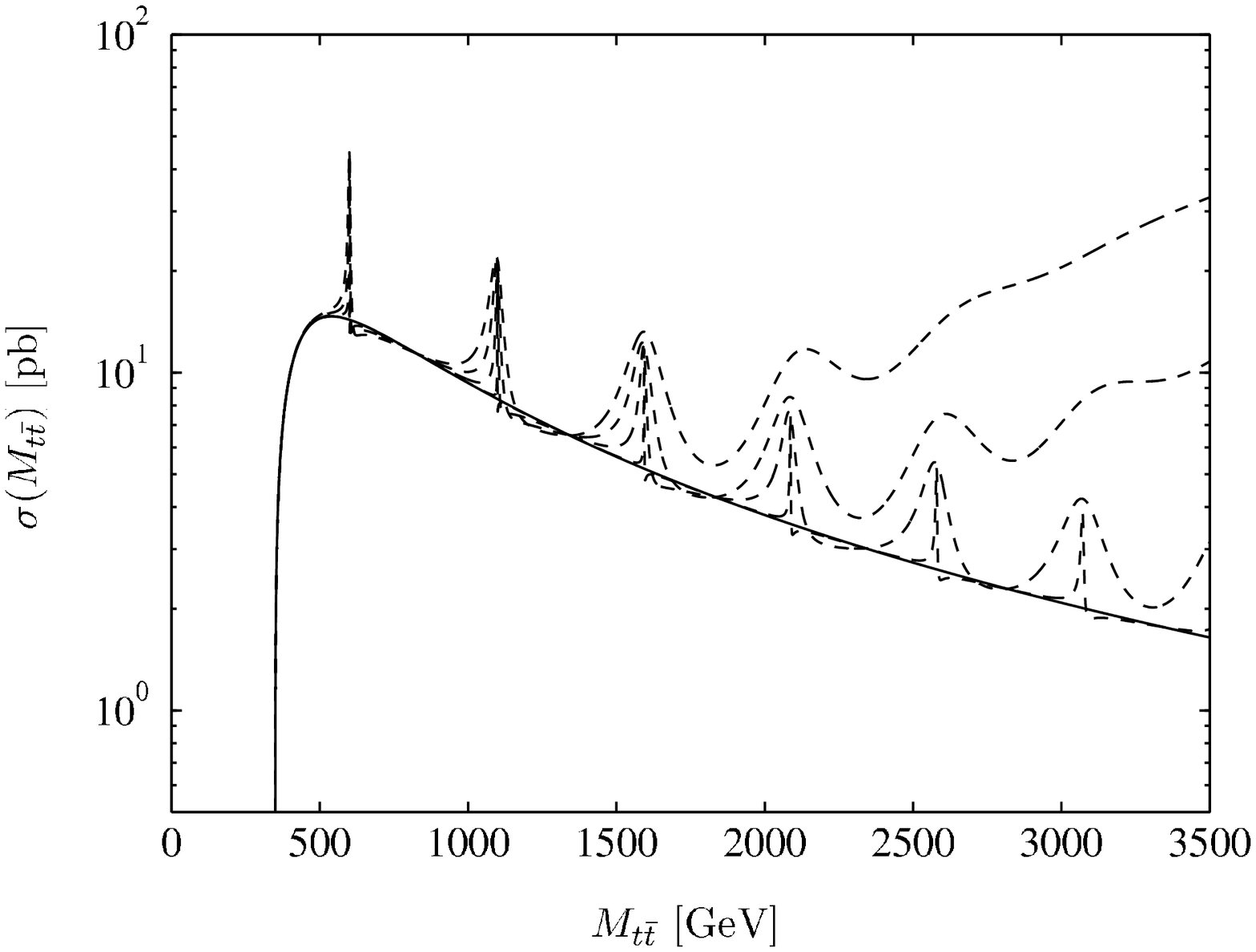}
\caption{
 The dependence of the 
 cross section of the top-antitop 
 quark pair production by gluon fusion 
 on the center-of-mass energy of colliding partons. 
The solid line and dashed lines correspond to 
 the results of the SM and the RS model for $m_1=600$ GeV 
 and $\kappa/{\bar{M}_{\rm pl}}=0.01, 0.04, 0.07$ 
 and $0.1$ from bottom to top, respectively.  
} 
\label{fig_sigma_omega_gg}
\end{center}
\end{figure}

In Figs.~\ref{fig_sigma_omega_qq} and \ref{fig_sigma_omega_gg}, 
 the 
 cross sections of the top-antitop pair production 
 through $q\bar{q}\rightarrow t\bar{t}$ (Fig.~\ref{fig_sigma_omega_qq})
 and $gg\rightarrow t\bar{t}$ (Fig.~\ref{fig_sigma_omega_gg}) 
 at the parton level are depicted as a function of 
 parton center-of-mass energy $\sqrt{s}=M_{t\bar{t}}$ 
 for $m_1=600$ GeV and various $\kappa/\bar{M}_{\rm pl}$. 
The SM cross section decreases, 
 while the cross section of the RS model grows rapidly with $\sqrt{s}$ 
 and thus the unitarity will be violated at high energies. 
This behavior can be understood from the formulas 
 of the squared amplitudes Eqs. \p{qq1}-\p{gg2}. 
Peaks in the figures correspond to resonant productions 
 of KK gravitons. 
Total cross sections and the width of each peak become larger, 
 as $\kappa/\bar{M}_{\rm pl}$ is taken to be large.

\begin{figure}[H]
\begin{center}
  \epsfxsize=8.2cm
  \epsfbox{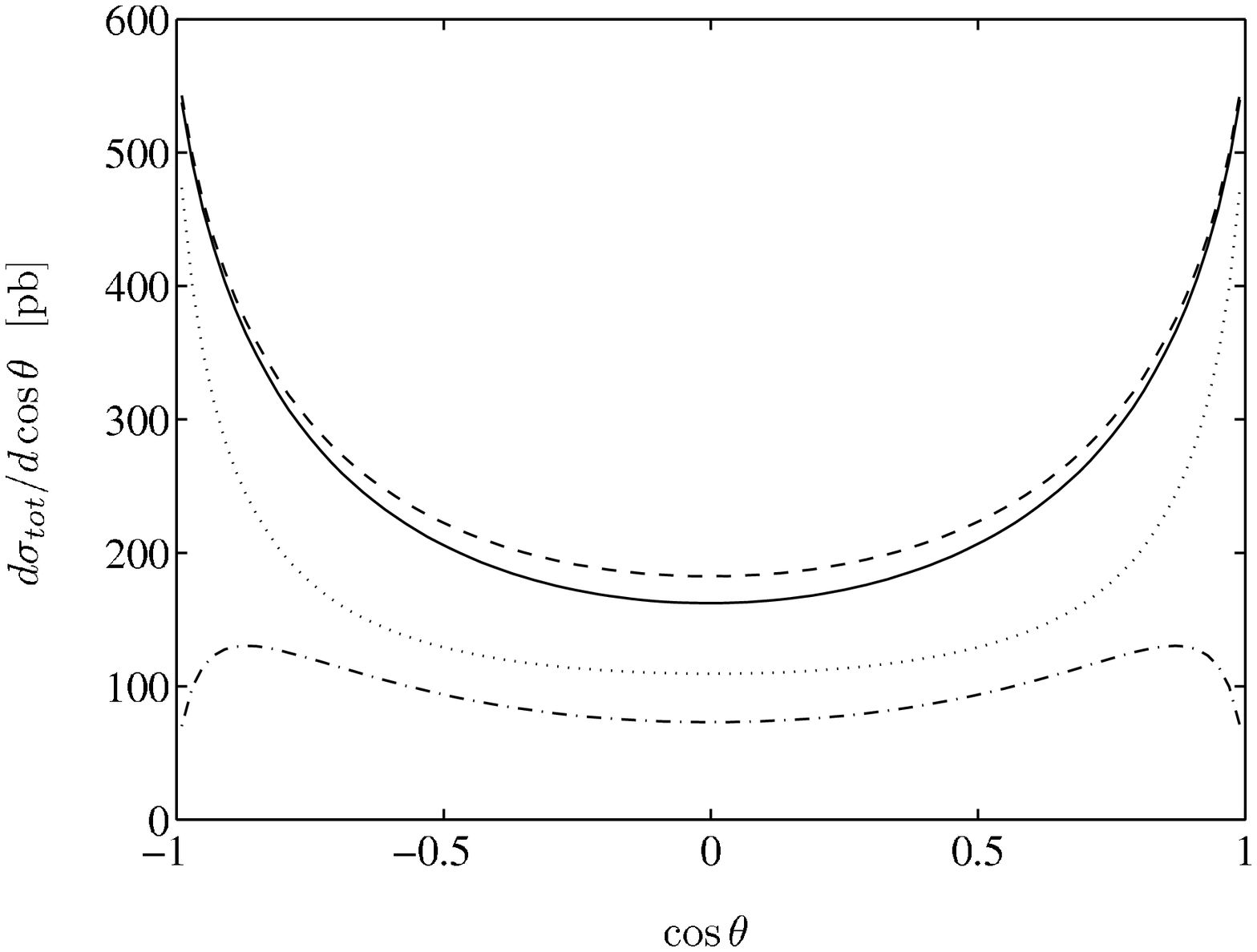}
\caption{
 Differential cross section \p{total_cosTheta} 
 as a function of $\cos\theta$ with $E_{CMS}=14$ TeV 
 for $m_1=600$ GeV and $\kappa/\bar{M}_{\rm pl}=0.1$. 
The solid line and dashed line correspond to 
 the results of the SM and the RS model, respectively. 
The differential cross sections 
 for the like (dotted) and the unlike (dash-dotted) 
 top-antitop spin pair productions 
 in the RS model are also depicted. 
}
\label{fig_cosTheta}
\end{center}
\end{figure}

Differential cross section for the top-antitop pair production 
 given by 
\begin{eqnarray}
 \frac{d \sigma_{tot}(pp \rightarrow t\bar{t})}{d\cos\theta}=
  \sum_{a,b} \int d x_1 \int d x_2\, 
   f_a(x_1,Q^2)f_b(x_2,Q^2)
   {d\sigma(t\bar{t}) \over d\cos \theta} 
   \label{total_cosTheta}
\end{eqnarray}
 is shown in Fig.~\ref{fig_cosTheta} with $E_{CMS}=14$ TeV. 
Here, the decomposition of the total cross section 
 into the like 
 ($t_{\uparrow}\bar{t}_{\uparrow}
  +t_{\downarrow}\bar{t}_{\downarrow}$) 
 and the unlike 
 ($t_{\uparrow}\bar{t}_{\downarrow}
 +t_{\downarrow}\bar{t}_{\uparrow}$) 
 top-antitop spin pairs in the RS model are also shown.

\begin{figure}[H]
\begin{center}
  \epsfxsize=8.2cm
  \epsfbox{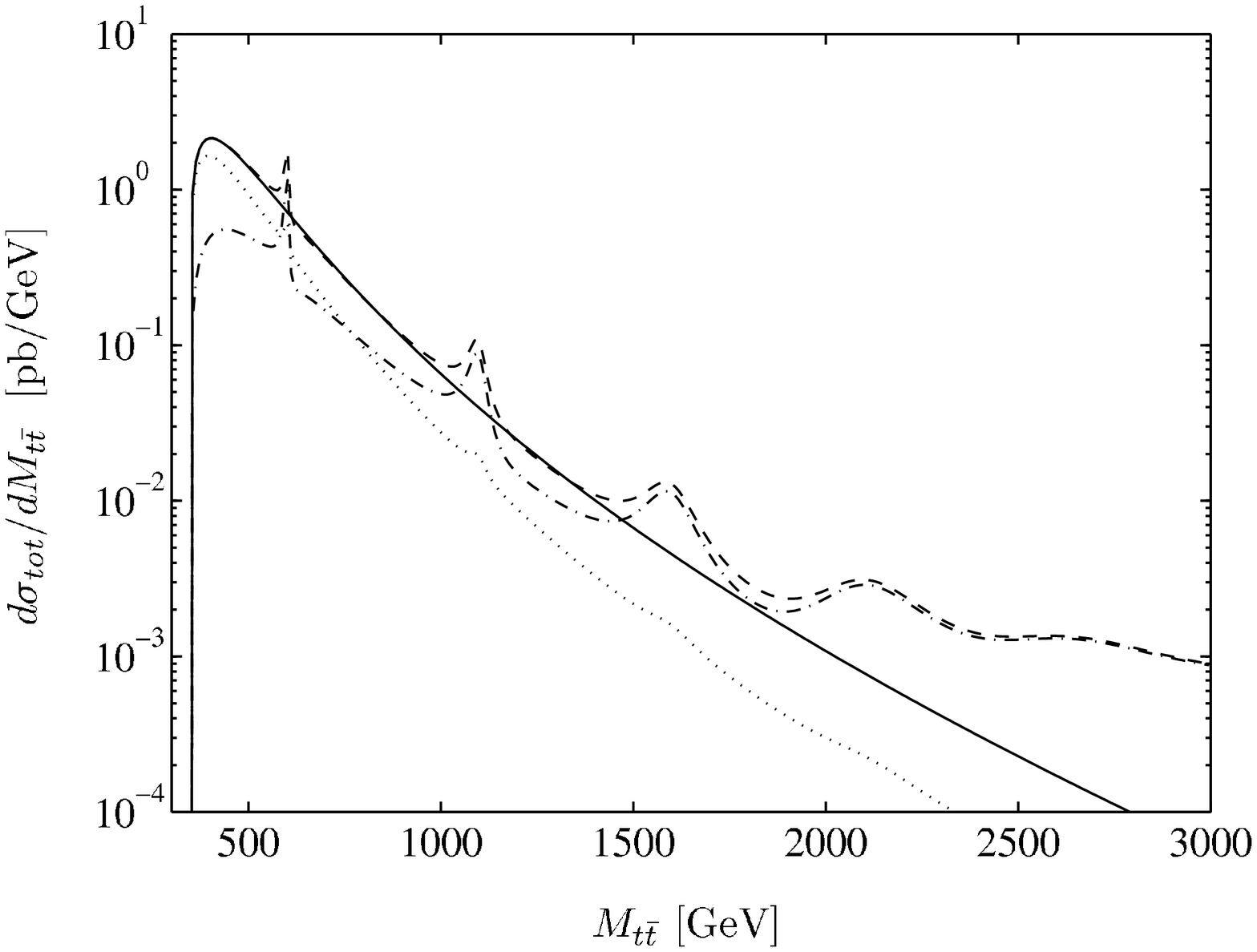}
\caption{
Differential cross section \p{total_s}
 as a function of the top-antitop invariant mass $M_{t\bar{t}}$ 
 for $m_1=600$ GeV and $\kappa/\bar{M}_{\rm pl}=0.1$. 
The solid and dashed lines correspond to 
 the results of the SM and the RS model, respectively. 
The differential cross sections 
 for the like (dotted) and the unlike (dash-dotted) 
 top-antitop spin pair productions in the RS model 
 are also depicted. 
}
\label{fig_sigma}
\end{center}
\end{figure}

\begin{figure}[H]
\begin{center}
  \epsfxsize=8.2cm
  \epsfbox{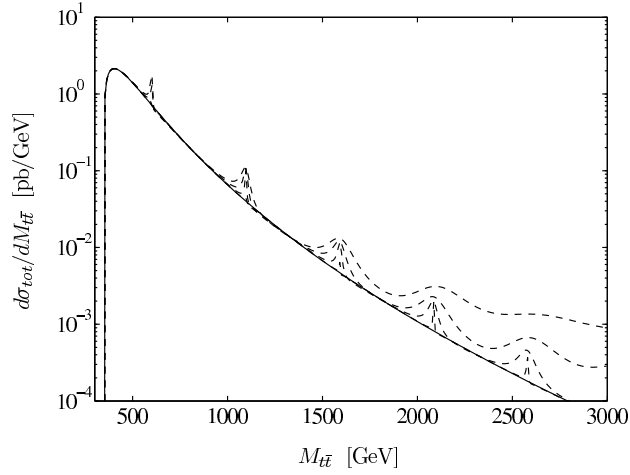}
\caption{
 Differential cross section \p{total_s} as a function 
 of the top-antitop invariant mass $M_{t\bar{t}}$. 
The solid and dashed lines correspond to 
 the results of the SM and the RS model 
 with $\kappa / \bar{M}_{\rm pl}=0.01, 0.04, 0.07$ and $0.1$ 
 from bottom to top, respectively.
} 
\label{fig_sigma_omega}
\end{center}
\end{figure}

We are also interested in the dependence of the cross section on
 the top-antitop invariant mass $M_{t\bar{t}}=\sqrt{(p_t+p_{\bar{t}})^2}$
 where $p_t(p_{\bar{t}})$ is momentum of (anti)top quark.
This is given by 
\begin{eqnarray}
 \frac{d \sigma_{tot}(pp \rightarrow t\bar{t})}{dM_{t\bar{t}}}=
  \sum_{a,b} \int\limits_{-1}^{1} d \cos\theta
   \int\limits_{\frac{M_{t\bar{t}}^2}{E_{CMS}^2}}^1 d x_1 
   \frac{2M_{t\bar{t}}}{x_1 E_{CMS}^2} f_a(x_1,Q^2)
   f_b\left(\frac{M_{t\bar{t}}^2}{x_1 E_{CMS}^2},Q^2\right)
   {d\sigma(t\bar{t}) \over d\cos \theta}.
   \label{total_s}
\end{eqnarray}
The result for $m_1=600$ GeV and $\kappa/\bar{M}_{\rm pl}=0.1$ 
 is shown in Fig.~\ref{fig_sigma}. 
The deviation of the cross section in the RS model  
 from the one in the SM grows as $s$ becomes large. 
Cross sections for the like and the unlike top-antitop spin pairs 
 in the RS model are also shown. 
The differential cross section as a function of 
 the center-of-mass energy of colliding partons 
 for various $\kappa/\bar{M}_{\rm pl}$ 
 is depicted in Fig.~\ref{fig_sigma_omega}.  
Deviation from the SM one becomes large 
 according to $\kappa/\bar{M}_{\rm pl}$.

\begin{figure}[H]
\begin{center}
  \epsfxsize=8.2cm
  \epsfbox{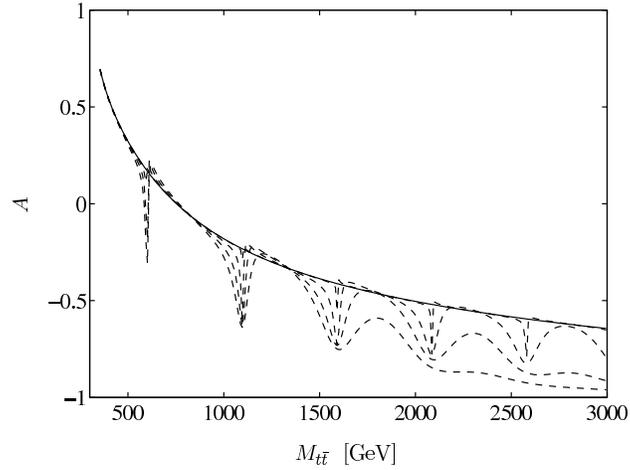}
\caption{
Spin asymmetry $\cal{A}$ as a function of 
 the top-antitop invariant mass $M_{t\bar{t}}$. 
The solid line corresponds to the SM, 
 while the dashed lines correspond to the RS model 
 with $\kappa /\bar{M}_{\rm pl}=0.01, 0.04, 0.07$ and $0.1$ 
 from up to down, respectively. 
}
\label{fig_A_omega}
\end{center}
\end{figure}

\begin{figure}[h]
\begin{center}
  \epsfxsize=8.2cm
  \epsfbox{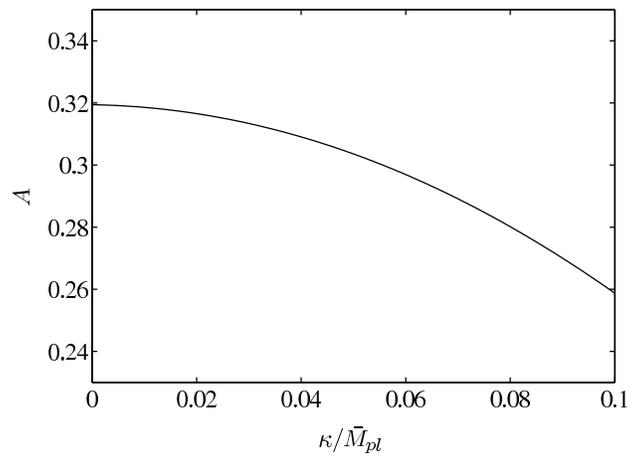}
\caption{
Spin asymmetry $\cal{A}$ as a function of 
 $\kappa / \bar{M}_{\rm pl}$ 
 at the LHC with $E_{CMS}=14$ TeV. 
As $\kappa/\bar{M}_{\rm pl}\rightarrow 0$, ${\cal A}$ becomes the SM
 value, $0.319$. 
}
\label{fig_A_total}
\end{center}
\end{figure}

Now let us show the result for the spin asymmetry $\cal{A}$. 
In Fig.~\ref{fig_A_omega}, 
 the spin asymmetry as a function of 
 the center-of-mass energy of colliding partons 
 for various 
 $\kappa/\bar{M}_{\rm pl}$ is depicted. 
Deviation from the SM one becomes larger 
 as the center-of-mass energy and $\kappa/\bar{M}_{\rm pl}$ 
 become larger. 
As expected, deviation is enhanced 
 around the poles of KK graviton resonances. 
This implies that we can expect a big statistical advantage 
 for the study of the top spin correlations 
 when we analyze experimental data around a pole. 
This fact is a crucial difference from the ADD model, 
 where no resonance of KK gravitons can be seen. 
In Fig.~\ref{fig_A_total} we show the spin asymmetry 
 $\cal{A}$ at the LHC, as a function of $\kappa/\bar{M}_{\rm pl}$. 
We can see a sizable deviation from the SM one, 
 for example, ${\cal A}=0.260$  for $\kappa/\bar{M}_{\rm pl} =0.1$. 
\section{Conclusions} 

In the RS model, 
 we have studied the top-antitop pair production 
 and the top spin correlations at the LHC. 
In addition to the Standard Model processes, 
 there is a new contribution to the top-antitop pair production 
 process mediated by graviton Kaluza-Klein modes in the $s$-channel. 
We have computed the corresponding density matrix 
 for the top-antitop pair production including the new contribution. 
We have shown various numerical results for 
 the production cross sections and the top spin correlations 
 with input parameters $m_1$ and $\kappa/\bar{M}_{\rm pl}$ 
 in the RS model. 
We have found a sizable deviation of the top-antitop pair production 
 cross section and the top spin correlations 
 from those in the Standard Model. 
In particular, resonant productions of the Kaluza-Klein gravitons 
 give rise to a remarkable enhancement of such deviations. 
This is a crucial difference from the case in the ADD model.

%
%
\vspace{1.0cm}
\noindent{\Large \bf Acknowledgements} \\
The work of M.A. is supported by the bilateral program of Japan Society 
 for the Promotion of Science and Academy of Finland, 
 ``Scientist Exchanges''. 
The work of N.O. is supported in part by Scientific Grants 
 from the Ministry of Education and Science of Japan.

%
%
\begin{appendix}
 \sect{The coefficients $\Delta_n^{yy}$}\label{coef}
Explicit forms of the coefficients $\Delta_n^{yy}$ are listed below. 
\begin{eqnarray}
 \Delta_{n}^{\gamma\gamma}&=&{1 \over 5}\,,\\
 \Delta_{n}^{gg}&=&{8 \over 5}\,, \\
 \Delta_{n}^{WW}&=&{2 \over 5}\sqrt{1-4r_W}
  \left({13 \over 12}+{14 \over 3}r_W+4r_W^2\right)\,, \\
 \Delta_{n}^{ZZ}&=&{1 \over 5}\sqrt{1-4r_Z}
  \left({13 \over 12}+{14 \over 3}r_Z+4r_Z^2\right)\,, \\
 \Delta_{n}^{HH}&=&{1 \over 30}(1-4r_H)^{5/2}\,, \\ 
 \Delta_{n}^{\nu\bar{\nu}}&=&{1 \over 10}\,, \\
 \Delta_{n}^{l\bar{l}}&=&{1 \over 10}(1-4r_l)^{3/2}
             \left(1+{8 \over 3}r_l\right)\,, \\
 \Delta_{n}^{q\bar{q}}&=&{3 \over 10}(1-4r_q)^{3/2}\left(1+{8 \over 3}r_l\right)\,,
\end{eqnarray}
where $r_y=m_y^2/m_n^2$ for each SM particle $y$. 
Expressions for leptons and quarks are for one flavor. 

\sect{Density matrix $R_{\alpha\beta,\alpha^\prime\beta^\prime}^i$}\label{matrix}
In this appendix, we give the full representation of the density 
 matrix for the top antitop pair production in the RS model.
For the $q\bar{q}\rightarrow t\bar{t}$ process, 
 we find
\begin{eqnarray}
R_{\uparrow\uparrow,\uparrow\uparrow}^q&=&
 R_{\downarrow\downarrow,\downarrow\downarrow}^q=
 -R_{\uparrow\uparrow,\downarrow\downarrow}^q=
 -R_{\downarrow\downarrow,\uparrow\uparrow}^q \nonumber \\ 
 &=&{g_s^4 \over 9}(1-\beta^2)\sin^2\theta
  +{s^4\beta^2 \over 128}|A(s)|^2(1-\beta^2)\sin^2 2\theta, \\
R_{\uparrow\uparrow,\uparrow\downarrow}^q&=&
 R_{\uparrow\uparrow,\downarrow\uparrow}^q=
 R_{\uparrow\downarrow,\uparrow\uparrow}^q=
 R_{\downarrow\uparrow,\uparrow\uparrow}^q=  
 -R_{\uparrow\downarrow,\downarrow\downarrow}^q=
 -R_{\downarrow\uparrow,\downarrow\downarrow}^q=
 -R_{\downarrow\downarrow,\uparrow\downarrow}^q=
 -R_{\downarrow\downarrow,\downarrow\uparrow}^q \nonumber \\            
 &=&{g_s^4 \over 9}\sqrt{1-\beta^2}\sin\theta\cos\theta
 -{s^4\beta^2 \over 32}{\rm Re}(A(s))\sqrt{1-\beta^2}\sin 4\theta, \\
R_{\uparrow\downarrow,\uparrow\downarrow}^q&=&
 R_{\downarrow\uparrow,\downarrow\uparrow}^q
 ={g_s^4 \over 9}(1+\cos^2\theta)
  +{s^4\beta^2 \over 128}|A(s)|^2(\cos^22\theta+\cos^2\theta), \\
R_{\uparrow\downarrow,\downarrow\uparrow}^q&=&
 R_{\downarrow\uparrow,\uparrow\downarrow}^q
 ={g_s^4 \over 9}(-1+\cos^2\theta)
  +{s^4\beta^2 \over 16}{\rm Re}(A(s))(1+2\cos 2\theta)\sin^2\theta,
\end{eqnarray}
and, for the $gg\rightarrow t\bar{t}$ process, 
\begin{eqnarray}
R_{\uparrow\uparrow,\uparrow\uparrow}^g&=&
 R_{\downarrow\downarrow,\downarrow\downarrow}^g \nonumber \\
 &=&\left({g_s^4 \over 96}{\cal Y}(\beta,\cos\theta)
  (1+\beta^2+\beta^2\sin^4\theta)
  +{\cal Z}(\beta,\theta,s)s^2 \sin^4\theta\right)
  \beta^2(1-\beta^2), \\
R_{\uparrow\uparrow,\downarrow\downarrow}^g&=&
 R_{\downarrow\downarrow,\uparrow\uparrow}^g \nonumber \\
 &=&
  -\left({g_s^4 \over 96}{\cal Y}(\beta,\cos\theta)
   (-1+\beta^2+\beta^2\sin^4\theta)
   +{\cal Z}(\beta,\theta,s)s^2 \sin^4\theta\right)
   \beta^2(1-\beta^2), \\
R_{\uparrow\uparrow,\uparrow\downarrow}^g&=&
 R_{\uparrow\uparrow,\downarrow\uparrow}^g=
 R_{\uparrow\downarrow,\uparrow\uparrow}^g=
 R_{\downarrow\uparrow,\uparrow\uparrow}^g=  
 -R_{\uparrow\downarrow,\downarrow\downarrow}^g=
 -R_{\downarrow\uparrow,\downarrow\downarrow}^g=
 -R_{\downarrow\downarrow,\uparrow\downarrow}^g=
 -R_{\downarrow\downarrow,\downarrow\uparrow}^g \nonumber \\            
 &=& \left({g_s^4 \over 96}{\cal Y}(\beta,\cos\theta)
  \cos\theta
  +{\cal Z}(\beta,\theta,s)s^2\right)\beta^2\sqrt{1-\beta^2}\cos\theta
   \sin^3 \theta, \\
R_{\uparrow\downarrow,\uparrow\downarrow}^g&=&
 R_{\downarrow\uparrow,\downarrow\uparrow}^g
 =\left({g_s^4 \over 96}{\cal Y}(\beta,\cos\theta)
  +{\cal Z}(\beta,\theta,s)s^2 
  \right) \beta^2(1+\cos^2\theta)\sin^2\theta , \\
R_{\uparrow\downarrow,\downarrow\uparrow}^g&=&
 R_{\downarrow\uparrow,\uparrow\downarrow}^g
 =-\left({g_s^4 \over 96}{\cal Y}(\beta,\cos\theta)
  +{\cal Z}(\beta,\theta,s)s^2 \right)\beta^2\sin^4\theta,
\end{eqnarray}
where the functions ${\cal Y}(\beta,\cos\theta)$ 
 and ${\cal Z}(\beta,\theta,s)$ are defined in \p{funcs1} and \p{funcs2}.

\end{appendix}
%
%

\newpage


\end{document}